\documentclass[10pt]{asme2ej}

\usepackage{epsfig} 
\usepackage{graphics}
\usepackage[colorlinks]{hyperref}
\usepackage[capitalize,nameinlink]{cleveref}
\usepackage[numbers]{natbib}
\title{Comparison of Random Forest and Neural Network Framework for Prediction of Fatigue Crack Growth Rate in Nickel Superalloys}

\author{Raghunandan Pratoori
    \affiliation{
	Department of Aerospace Engineering\\
    Email: rnp@iastate.edu
    }
}

\begin{document}

\maketitle

\begin{abstract}
{\it The rate of fatigue crack growth in Nickle superalloys is a critical factor of safety in the aerospace industry. A machine learning approach is chosen to predict the fatigue crack growth rate as a function of the material composition, material properties and environmental conditions. Random forests and neural network frameworks are used to develop two different models and compare the two results. Both the frameworks give good predictions with $r^2$ of 0.9687 for random forest and 0.9831 for neural network.

}
\end{abstract}


\section{Introduction}

Superalloys are of great utility in the fields of engineering, especially in aerospace and power generation industries because of their excellent balance of mechanical and chemical properties~\cite{Pollock2006}. These superalloy components are subjected to high operating stresses under cyclic loading. This makes fatigue behavior a very important factor to be considered. Fatigue is caused due to fluctuations in the mechanical and thermal loads in various stages of operation. The fatigue crack propagation is dependent on the various factors like environmental conditions, microstructure, composition of the material and the like~\cite{Bathias1999, Murakami2000, Shyam2004, Bhadeshia1999}. This makes a case for developing a model to predict the fatigue crack growth rate under various conditions.

Understanding the complexity of the phenomena, in this work, a machine learning approach is chosen to develop a model for the prediction of fatigue crack growth rate. Machine learning approaches are preferred in situation where calculation of some attributes exactly in real-time is not possible. Among the various available machine learning approaches, random forest and neural network frameworks are chosen. In the recent past, many researchers have used random forests in the field of material science to calculate properties of materials which need extreme experimental conditions to determine them, to establish a relationship between two or more material properties or to enhance the material design process.

Random forest has successfully been used to enhance the prediction accuracy in the fields of ecology~\cite{Grimm2008,Prasad2006}, remote sensing~\cite{Gislason2006, Ham2005, Lawrence2006, Pal2005}, material science~\cite{Carrete2014, nagasawa2018computer, vinci2018understanding}. \citeauthor{Carrete2014} used random forests to identify the compounds with low lattice thermal conductivity and the critical properties influencing the lattice thermal conductivity. With this approach, approximately 79,000 half-Heusler entries in AFLOWLIB.org database were scanned through with much ease, which would have been a high cost and time-consuming experimental challenge. \citeauthor{nagasawa2018computer} have used random forest to design organic photovoltaic materials and demonstrated its utility in the synthesis and characterization of the polymer. \citeauthor{vinci2018understanding} have used random forest to investigate the mechanical properties of Ultra-High-Temperature-Ceramic-Matrix-Composites and also studied the influence of different parameters on different properties.

Neural networks are one of the popular machine learning techniques often used in a lot of applications such as pattern recognition~\cite{Nakamoto1990, Poulton1992}, image processing~\cite{Parisi1998, Suzuki2006}, biotechnology~\cite{Chen2015, Karim1997, Montague1994}, material sciences~\cite{Bhadeshia1999,  butler2018machine, Kotkunde2014, hassan2009prediction, singh2011prediction}, etc. The accurate results obtained from it and the simple representation makes it applicable to almost all the areas of research. \citeauthor{Kotkunde2014} have developed a neural network model enhanced with differential evolution algorithm to predict the flow stress values for Ti-6Al-4V alloy as a function of strain, strain rate and temperature. \citeauthor{hassan2009prediction} have used neural networks in predicting the physical properties like density and porosity and hardness of aluminium–copper/silicon carbide composites as a function of weight percentage of copper and volume fraction of the reinforced particles. \citeauthor{singh2011prediction} have developed a neural network model for predicting the effective thermal conductivity of porous systems filled with different liquids.

\section{Methods}

In this work, two different frameworks - random forest and neural network, are tried to predict the fatigue crack growth rate.

\subsection{Random Forest}

\begin{figure}[htpb]
	\begin{center}
		\resizebox{140mm}{!}{\includegraphics[trim={60mm 50mm 60mm 50mm},clip]{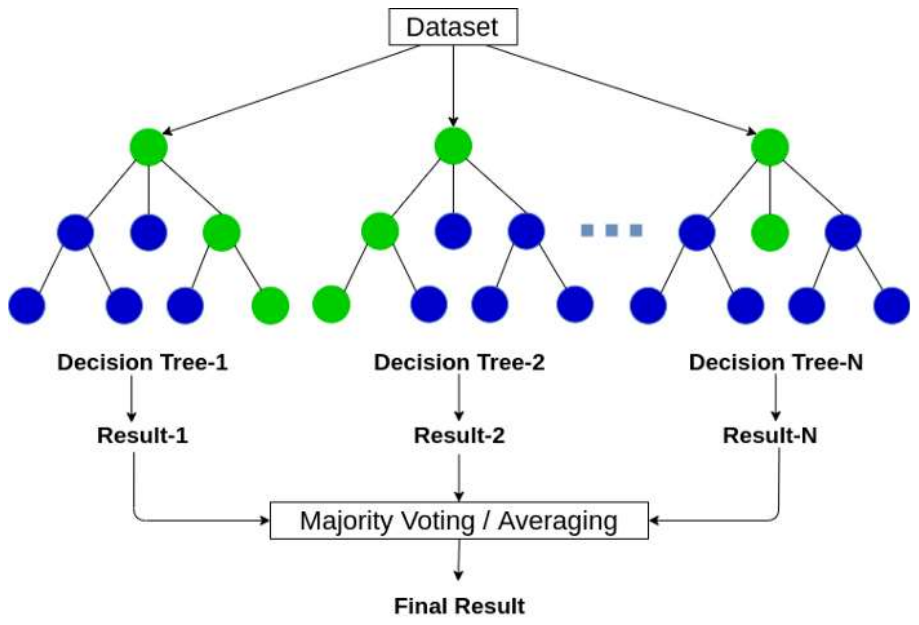}}
		\caption{Pictorial representation of algorithm of random forest}
		\label{fig:algo_RF}
	\end{center}
\end{figure}

Random Forest is an ensemble of $n$ trees dependent on a $p$-dimensional vector of variables. The ensemble produces $n$ outputs one for each tree, which are averaged to produce one final prediction (\Cref{fig:algo_RF}).

The training algorithm proceeds as follows:
\begin{enumerate}
    \item From the training data, draw a random sample with replacement (bootstrapping).
    \item For each bootstrap sample, grow a tree choosing the best split among a randomly selected subset of variables until no further splits are possible.
    \item Repeat the above steps until $n$ such trees are grown.
\end{enumerate}

Cross-validation is in-built in the training step of random forest with the use of Out-Of-Bag (OOB) samples~\cite{breiman1996out}.So, We can calculate an ensemble prediction $Y^{OOB}$ by averaging only its OOB predictions. Calculate an estimate of the mean square error (MSE) for regression by
\begin{equation}
    MSE = n^{-1}\Sigma^n_{i=1}\{Y^{OOB}(X_i)-Y_i\}^2
\end{equation}

Random Forest algorithm is capable of selecting important variables but does not produce an explicit relationship between variables and the predictions~\cite{breiman2002using}. However, a measure of how each variable contributes can be estimated by measuring the node purity in the course of training.

\subsection{Neural Network}

\begin{figure}[htpb]
	\begin{center}
		\resizebox{140mm}{!}{\includegraphics[trim={40mm 30mm 50mm 30mm},clip]{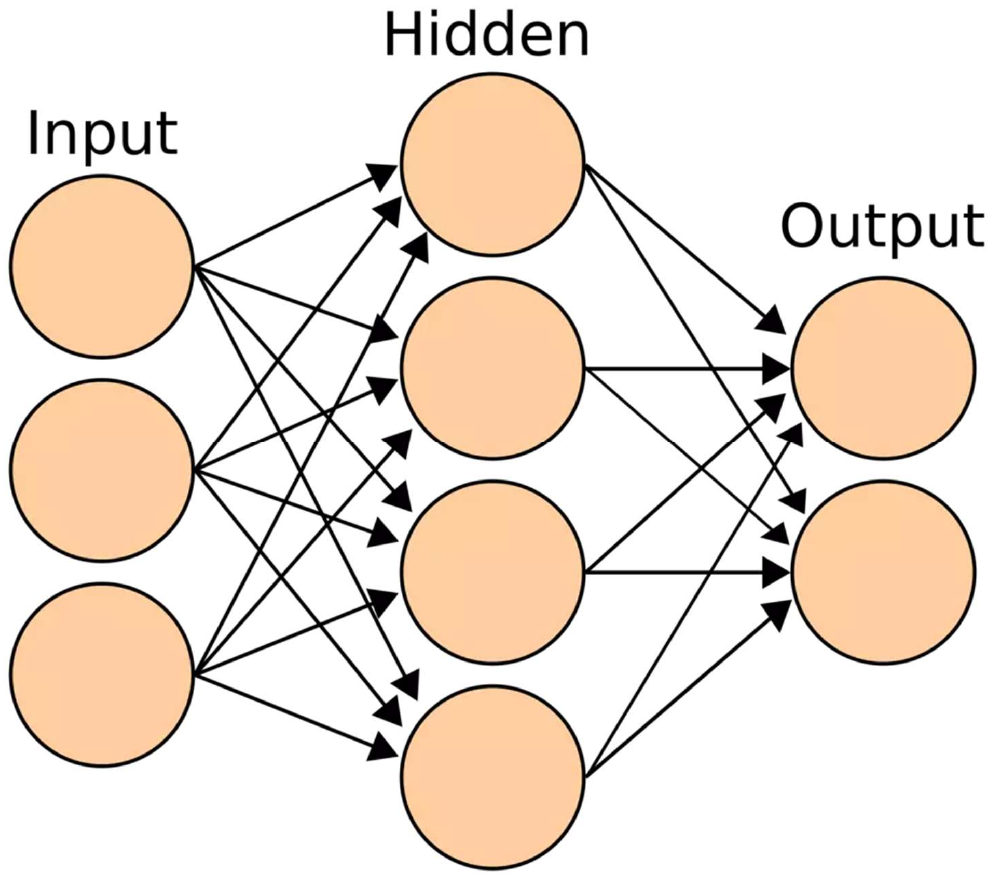}}
		\caption{Pictorial representation of algorithm of neural network}
		\label{fig:algo_NN}
	\end{center}
\end{figure}

A neural network is made of neurons which combine all the inputs given to the neuron and transfers it to another neuron based on an activation function. Tan sigmoid, linear line and Log sigmoid are in general the popularly
used activation functions or transfer functions. A group of neurons connecting together in a
weighted form to give rise to output is called a layer of neurons. The layout
of a single layer neural network is shown in \Cref{fig:algo_NN}. The weights of each layer are evaluated by using a back propagation algorithm. The transfer function relating the inputs and the hidden layer is given by
\begin{equation}
    h_i=\tanh (\Sigma_j w_{ij}^{(1)}x_j+\theta_i^{(1)})
\end{equation}
The relationship between the hidden units and the output is given by
\begin{equation}
    y=\Sigma_i w_{i}^{(2)}h_i+\theta^{(2)}
\end{equation}

\section{Data}

\begin{figure}[htpb]
	\begin{center}
		\resizebox{110mm}{!}{\includegraphics[angle=90,origin=c,trim={0mm 40mm 0mm 40mm},clip]{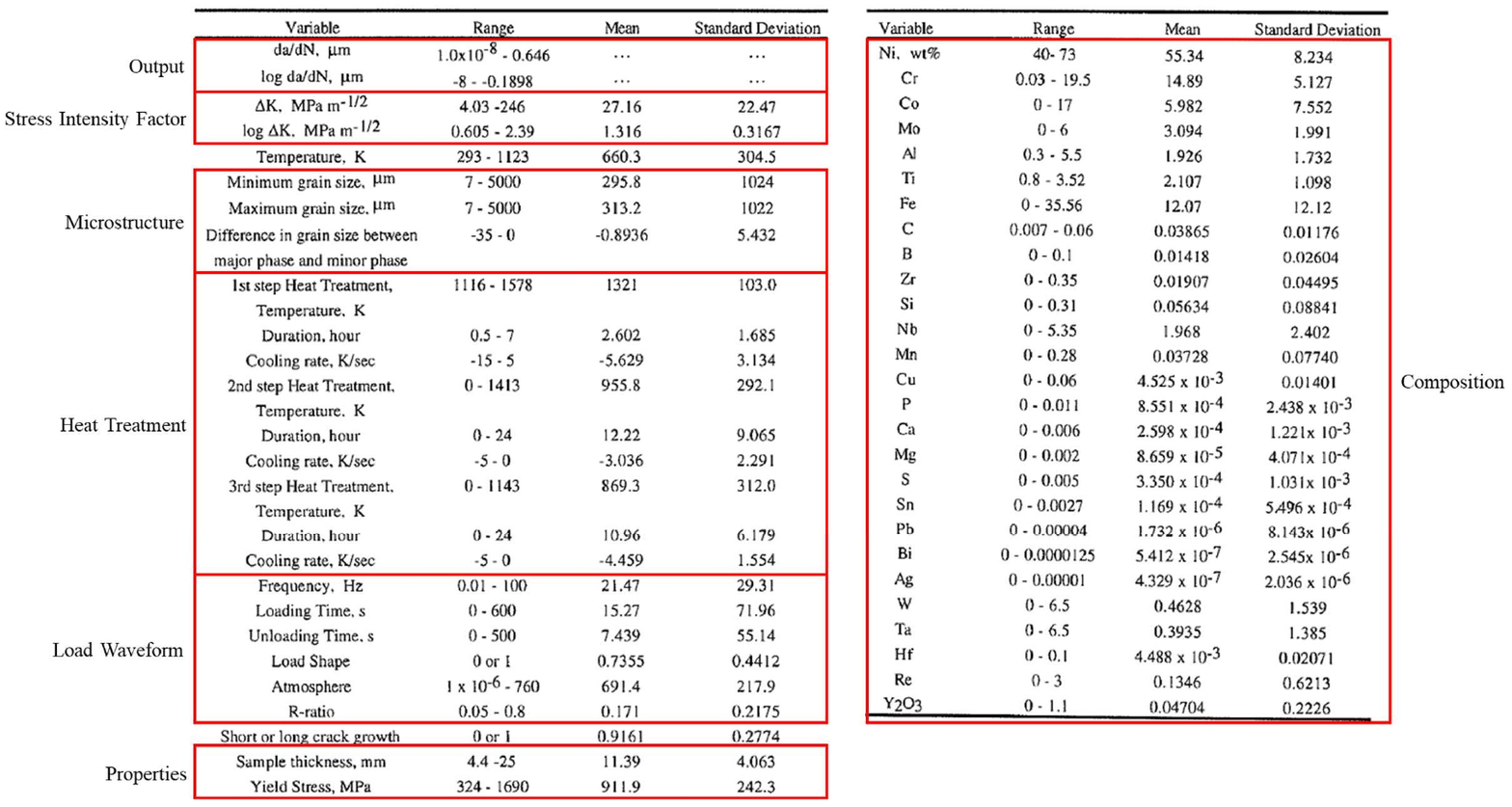}}
		\caption{Summary of the dataset used~\cite{FujllH.;MackayD.J.C;Bhadeshia1996}}
		\label{fig:data}
	\end{center}
\end{figure}

The data set used in this work is from published literature. This data set consists of 1894 data points for fatigue crack growth which is dependent on 51 input variables which can be categorized into stress intensity factor, temperature, microstructure, heat treatment, load waveform, type of crack growth, properties of the material and composition. The data is summarized as shown in \Cref{fig:data}. A detailed account of all the input variables is described in <Fuiji>. Type of crack growth is binary valued with short crack growth represented by 0 and long crack growth represented by 1. Since the problem here is modelled as a regression problem, type of crack growth is omitted in the analysis and hence only 50 variables are being used. Among the output variables, $\log(da/dN)$ is chosen instead of $da/dN$ to build a regression model, following the Paris' law~\cite{paris1963critical}.

\section{Analysis}

As can be observed from \Cref{fig:data}, the range of the data varies significantly between the variables. To prevent any adversarial effect in determination of influence of the variables, both the input variables and output variables are normalized within the range [0, 1] as follows:
\begin{equation}
    x_N=\frac{x-x_{min}}{x_{max}-x_{min}}
\end{equation}
where $x_N$ is the normalized value of $x$, $x_{min}$ is the minimum value of each variable and $x_{max}$ is the maximum value of each variable of the original data. The normalized data is then analysed using the R~\cite{R} software using a random forest and neural network framework. 75\% of the data is randomly selected for training the models and tested with the remaining 25\% of the data.

\subsection{Random Forest}

Random forest framework is developed using \verb+randomforest+ package~\cite{rrf}. All the 50 variables are used in the initial run, assuming a linear relationship between the input and the output variables. The maximum number of decision trees is chosen to be 1000 and the number of trees that give the least mean square error is selected for further analysis. Since there are a total of 50 variables, there is a chance of model overfitting the data and hence leading to a poor performance with the test data. To avoid this, the factors with low \%IncMSE and IncNodePurity are omitted and again analysed in the same fashion as the initial run.

\subsection{Neural Network}

Neural network framework is developed using \verb+neuralnet+ package. All the 50 variables are used in the initial run, assuming a linear relationship between the input and the output variables, just like in the case of random forest. In the initial run, only one hidden layer is considered with the number of nodes varying from 1 to 20. These models are trained on the training data and then applied on the testing data. The number of nodes with the highest $r^2$ is selected and a new layer is added to it and the number of nodes are varied from 1 to 20. Again, these are applied on the testing data and the number of nodes with the highest $r^2$ is selected. A sigmoid function is used as the activation function.

\section{Results}

\subsection{Random Forest}

\begin{figure}[htpb]
	\begin{center}
		\resizebox{140mm}{!}{\includegraphics{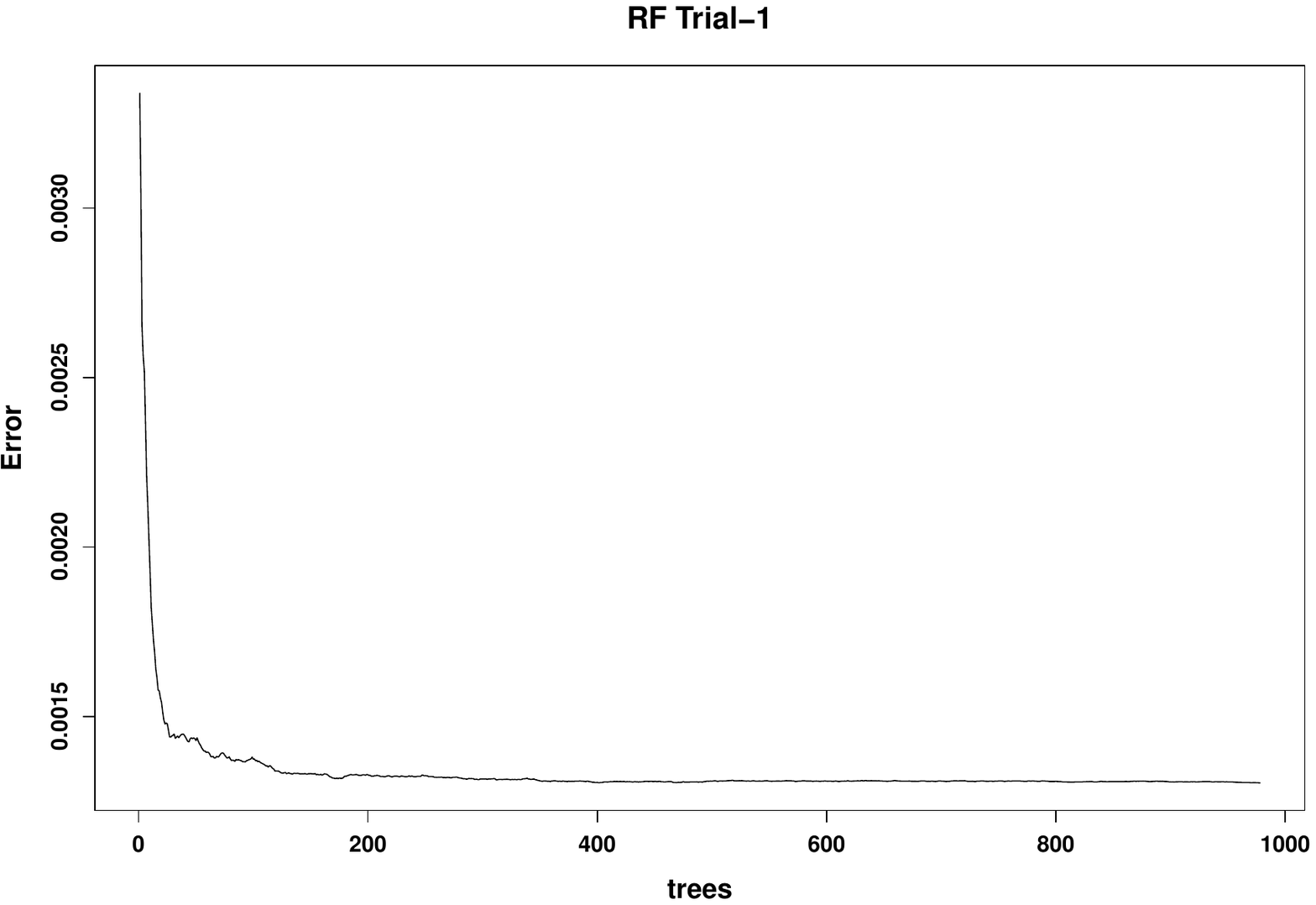}}
		\caption{Change in mean square error with increase in number of decision trees 50 variables}
		\label{fig:RFT1}
	\end{center}
\end{figure}

\begin{figure}[htpb]
	\begin{center}
		\resizebox{140mm}{!}{\includegraphics{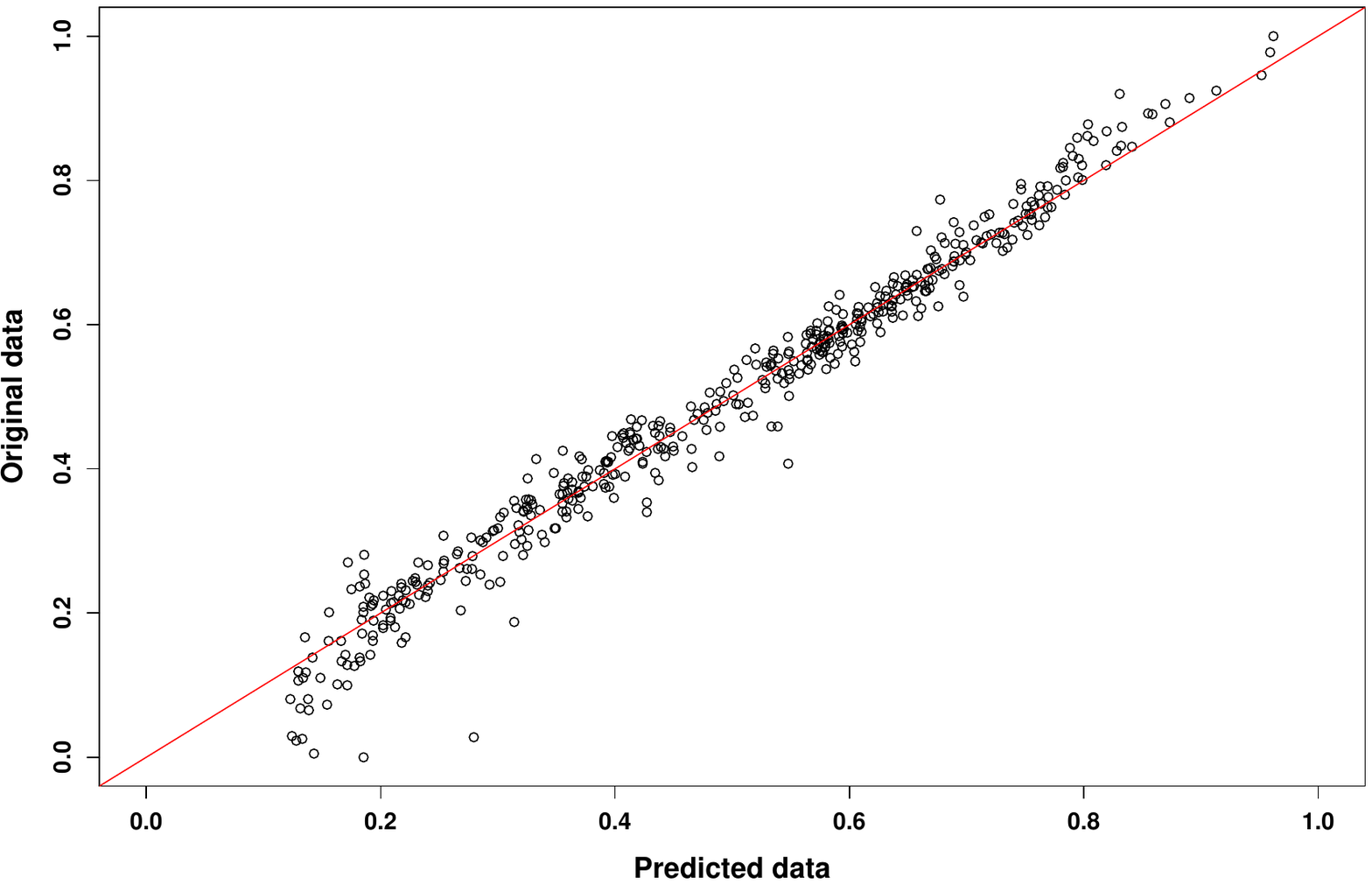}}
		\caption{Comparison of predicted $\log da/dN$ and the original data with random forest developed with 50 variables. The red line represents $x=y$.}
		\label{fig:com_RFT1}
	\end{center}
\end{figure}

\begin{figure}[htpb]
	\begin{center}
		\resizebox{140mm}{!}{\includegraphics{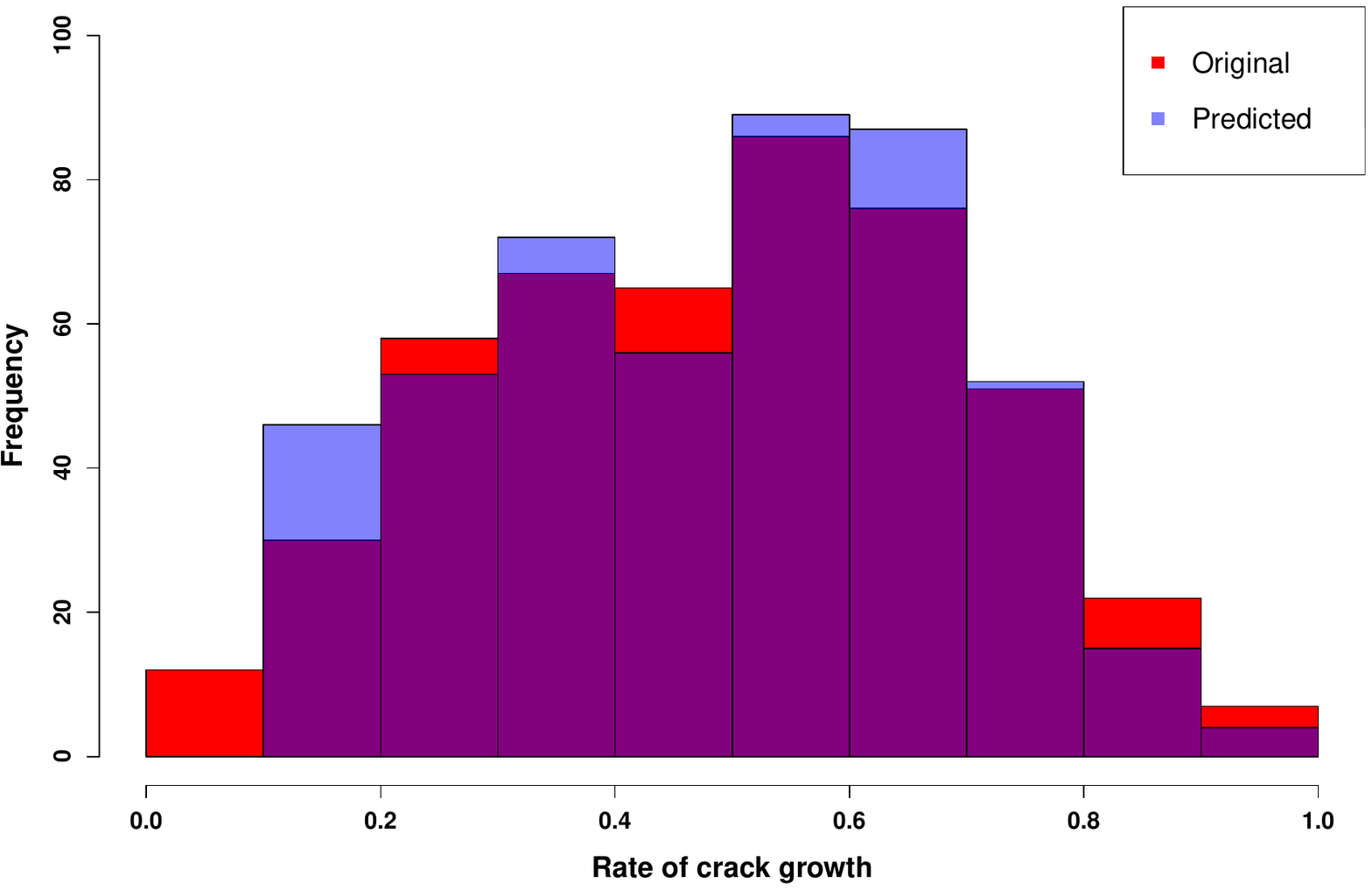}}
		\caption{Histogram comparing the distribution predicted and the original with random forest developed with 50 variables.}
		\label{fig:hist_RFT1}
	\end{center}
\end{figure}

For the first run with all the 50 variables, the change in mean square error with increase in number of decision trees is shown in \Cref{fig:RFT1}. The least error is observed with a choice of 978 decision trees and a random forest framework is designed with the same. The number of variables tried at each split are 16. The mean squared residual obtained from this model is $1.32\times 10^{-3}$. The percentage variance explained by this framework is $96.93\%$. This shows the model is a good fit for the training data. Using this framework on the testing data with and comparing with the original testing data, gives an $r^2$ value of 0.9693. The correlation value of 0.9865 and p-value of $<2.2\times10^{-16}$ show there is a significant correlation with the chosen variables and the data. \Cref{fig:com_RFT1} shows the comparison of the predicted values of $\log da/dN$ to the original values. It can be clearly seen from \Cref{fig:com_RFT1} that the points align close to the $x=y$ line with major deviations present near the lower end. This shows that the model is not a good representation when the rate of crack growth is smaller. \Cref{fig:hist_RFT1} shows the distribution of the original and the predicted rate of crack growth. It can be clearly observed that the original data in the range (0, 0.1) has not been predicted accurately and even the data in the range (0.8, 1.0) has not been completely reproduced.

\begin{table}[]
\caption{Mean Decrease Accuracy (\%IncMSE) and Mean Decrease Gini (IncNodePurity) of each variable in the initial run of  random forest}
\begin{center}
\begin{tabular}{|l|l|l|l|l|l|}
\hline
Variable                             & \%IncMSE  & IncNodePurity & Variable                   & \%IncMSE & IncNodePurity \\ \hline
Atm.Pressure                         & 72.645326 & 0.435396      & Cobalt..wt..               & 13.21943 & 0.582949      \\ \hline
delta.K..MPa.sqrt.m..                & 50.245341 & 12.06703      & Nickel..wt..               & 12.33514 & 0.26295       \\ \hline
Frequency..Hz.                       & 48.807527 & 15.70495      & HT2.Temp..K.               & 12.25707 & 1.019443      \\ \hline
log10.delta.K.                       & 47.736954 & 10.87632      & Titanium..wt..             & 12.24269 & 0.513162      \\ \hline
Yield.Strength..MPa.                 & 44.850812 & 0.503051      & Unloading.Time..s.         & 11.82461 & 0.286359      \\ \hline
Temp..K.                             & 35.091226 & 4.990744      & HT1.CoolingRate..      & 11.81688 & 0.057953      \\ \hline
Min.Grain.Size..micro.m.             & 29.230863 & 0.259286      & HT3.CoolingRate..      & 11.65292 & 0.027644      \\ \hline
Max.Grain.Size..micro.m.             & 28.139714 & 0.196492      & Niobium..wt..              & 11.3235  & 0.084579      \\ \hline
Loading.Time..s.                     & 27.477331 & 2.021161      & HT2.CoolingRate...      & 10.97752 & 0.06581       \\ \hline
HT1.Temp..K.                         & 26.732803 & 0.278611      & Manganese..wt..            & 10.45497 & 0.213812      \\ \hline
Diff.in.GS. & 26.287002 & 0.095541      & Copper..wt..               & 8.091316 & 0.031112      \\ \hline
Chromium..wt..                       & 26.195653 & 0.5403        & Phosphorus..wt..           & 7.709504 & 0.033819      \\ \hline
R.ratio                              & 23.602636 & 0.486236      & Sulphur..wt..              & 7.652882 & 0.020254      \\ \hline
Thickness..mm.                       & 19.991276 & 0.589908      & HT1.Time..hrs.             & 7.056421 & 0.160201      \\ \hline
HT3.Temp..K.                         & 19.269015 & 0.077698      & Tungsten..wt..             & 6.754946 & 0.01083       \\ \hline
HT3.Time..hrs.                       & 18.495807 & 0.14059       & Tantalum..wt..             & 6.463679 & 0.023136      \\ \hline
Molybdenum..wt..                     & 17.562076 & 2.840485      & Yttrium.Oxide..wt.. & 5.813943 & 0.035488      \\ \hline
Iron..wt..                           & 17.265415 & 0.076556      & Silver..wt..               & 4.174516 & 0.001748      \\ \hline
Silicon..wt..                        & 17.074731 & 0.123597      & Rhenium..wt..              & 4.118154 & 0.00736       \\ \hline
HT2.Time..hrs.                       & 16.238195 & 0.07077       & Lead..wt..                 & 4.059671 & 0.003141      \\ \hline
Load.Shape                           & 16.116972 & 0.546295      & Magnesium..wt..            & 3.879866 & 0.002424      \\ \hline
Boron..wt..                          & 15.479676 & 2.378501      & Calcium..wt..              & 3.761607 & 0.00224       \\ \hline
Zirconium..wt..                      & 15.323161 & 0.382954      & Tin..wt..                  & 3.729041 & 0.001958      \\ \hline
Carbon..wt..                         & 14.446973 & 0.210651      & Hafmium..wt..              & 3.342515 & 0.006146      \\ \hline
Aluminium..wt..                      & 14.203788 & 0.13301       & Bismuth..wt..              & 3.321477 & 0.001853      \\ \hline
\end{tabular}
\end{center}

\label{tab:RFT1}
\end{table}

\begin{figure}[htpb]
	\begin{center}
		\resizebox{140mm}{!}{\includegraphics{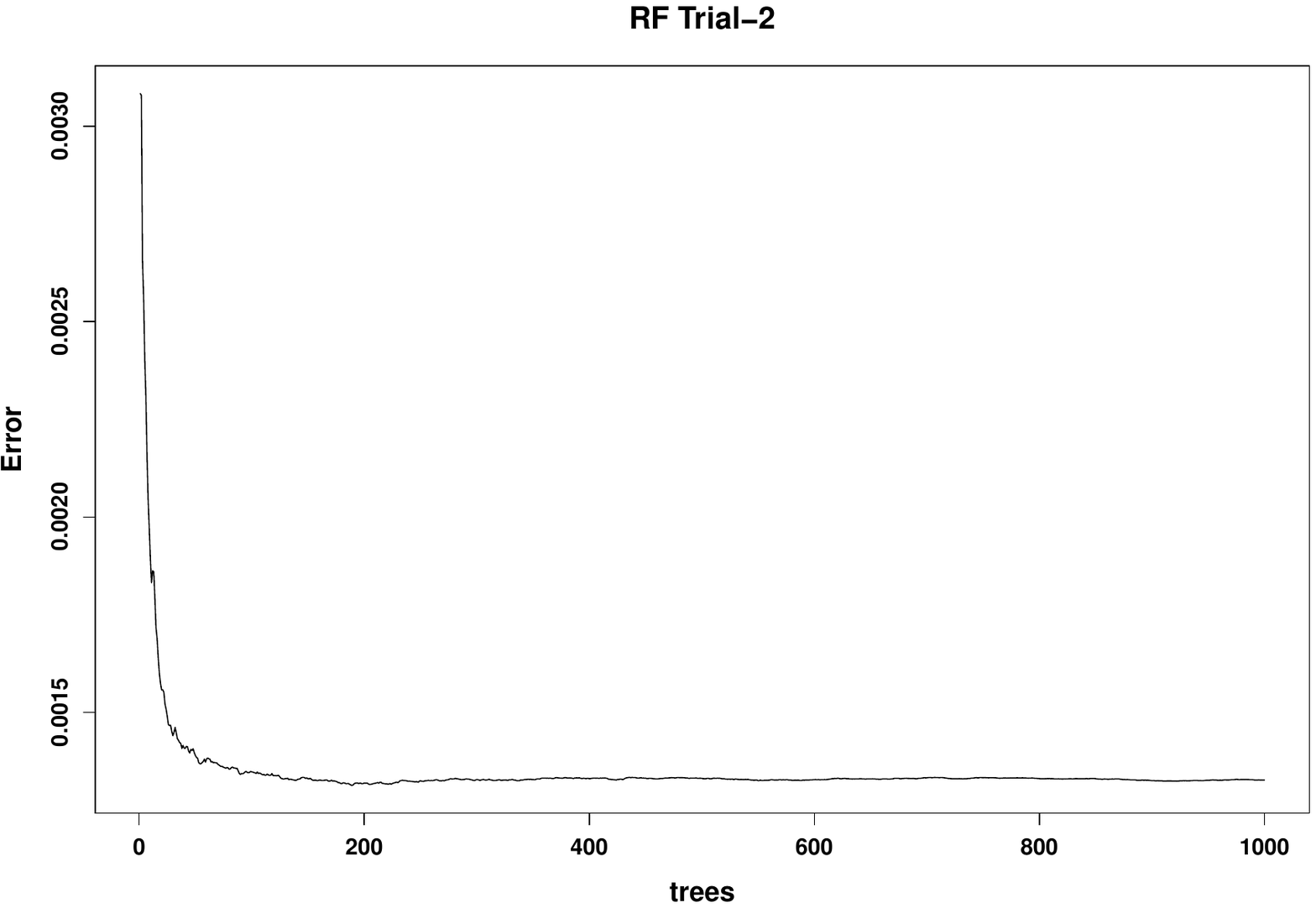}}
		\caption{Change in mean square error with increase in number of decision trees with 35 variables}
		\label{fig:RFT2}
	\end{center}
\end{figure}

\begin{figure}[htpb]
	\begin{center}
		\resizebox{140mm}{!}{\includegraphics{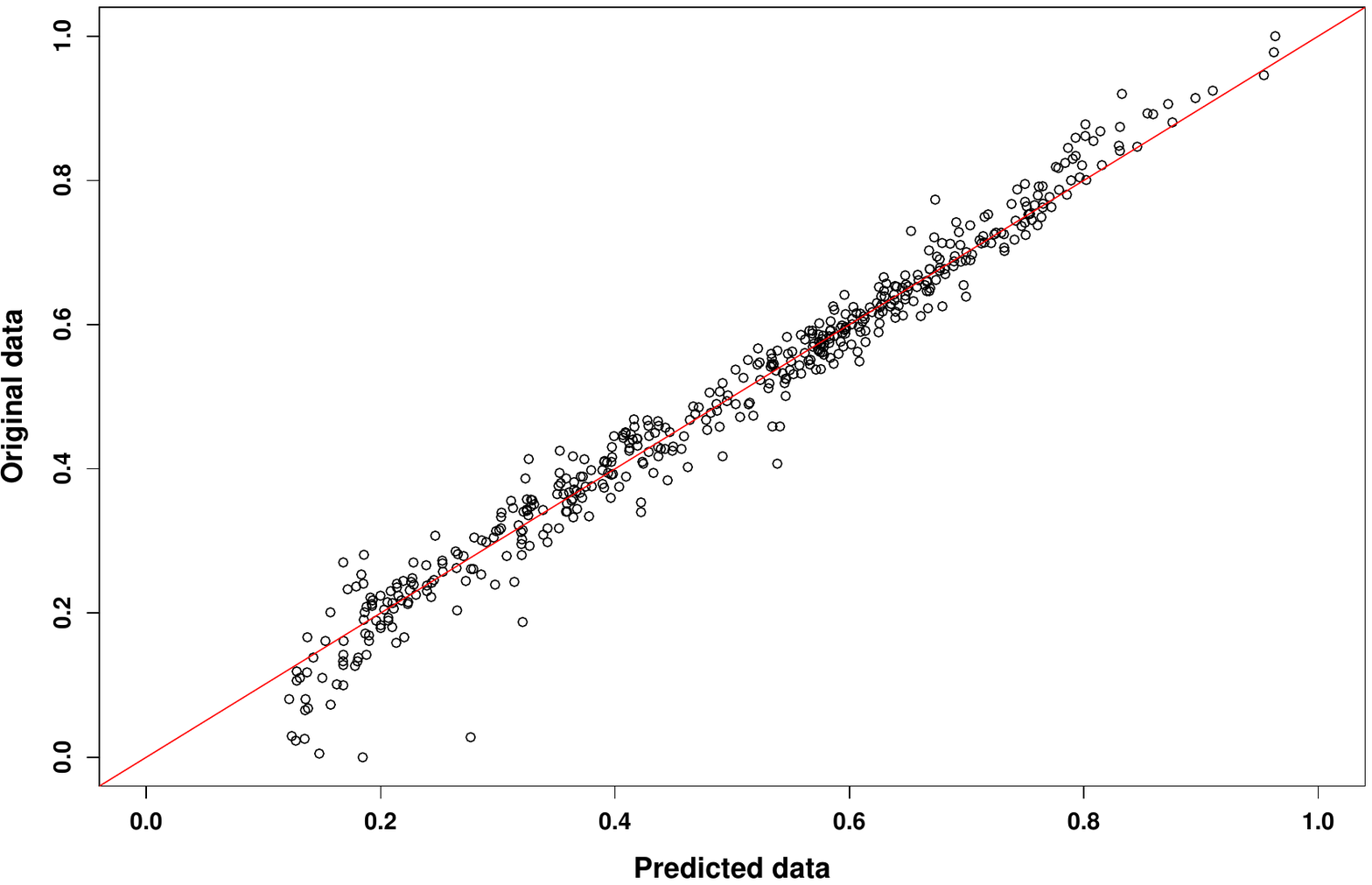}}
		\caption{Comparison of predicted $\log da/dN$ and the original data with random forest developed with 35 variables. The red line represents $x=y$.}
		\label{fig:com_RFT2}
	\end{center}
\end{figure}

\begin{figure}[htpb]
	\begin{center}
		\resizebox{140mm}{!}{\includegraphics{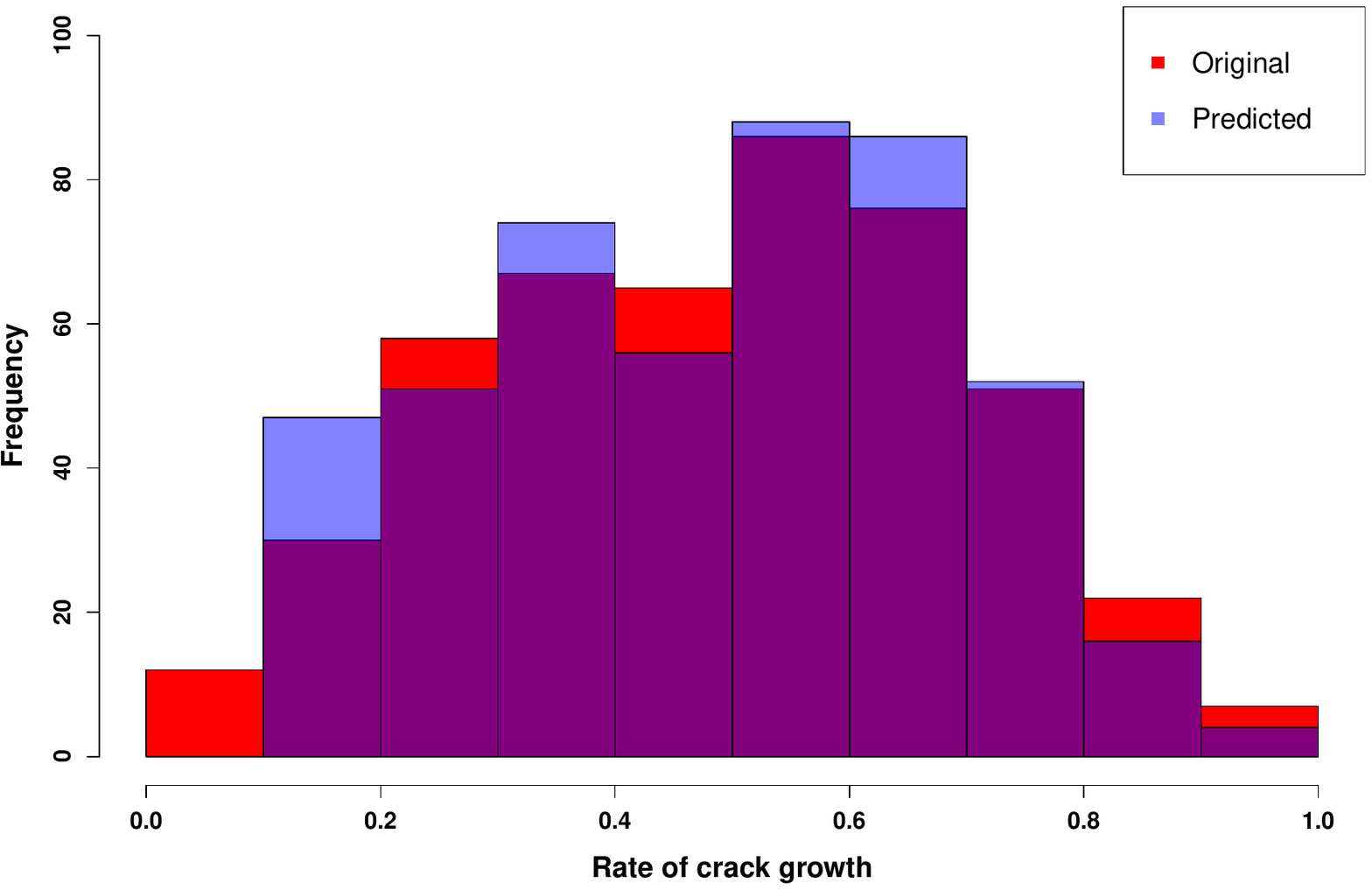}}
		\caption{Histogram comparing the distribution predicted and the original with random forest developed with 35 variables.}
		\label{fig:hist_RFT2}
	\end{center}
\end{figure}

\Cref{tab:RFT1} shows the values of Mean Decrease Accuracy and Mean Decrease Gini sorted according to decreasing Mean Decrease Accuracy. The variables with low Mean Decrease Accuracy and Mean Decrease Gini are not significant and can be omitted. In this case, the variables with the Mean Decrease Accuracy less than 5 and Mean Decrease Gini less than 0.005 are omitted leaving us with 35 variables. For the final run, the same procedure is followed as that of the initial run. The change in mean square error with increase in number of decision trees is shown in \Cref{fig:RFT2}. The least error is observed with a choice of 189 decision trees and a random forest framework is designed with the same. The number of variables tried at each split are 11. The mean squared residual obtained from this model is $1.31\times 10^{-3}$ and the percentage variance explained by this framework is $96.94\%$, which are very close to the values observed in the initial run. Using this framework on the testing data with and comparing with the original testing data, gives an $r^2$ value of 0.9687 which is 0.06\% lower than the initial run. The correlation value of 0.9862 and p-value of $<2.2\times10^{-16}$ show there is a significant correlation with the chosen variables and the data. Looking at the values of $r^2$ and correlation coefficient, it can be attested that the omitted values are insignificant in determining the rate of crack growth. \Cref{fig:com_RFT2} shows the comparison of the predicted values of $\log da/dN$ to the original values. It can be clearly seen from \Cref{fig:com_RFT2} that the points align close to the $x=y$ line with major deviations present near the lower end similar to the initial run. This shows that the model is not a good representation when the rate of crack growth is smaller. \Cref{fig:hist_RFT2} shows the distribution of the original and the predicted rate of crack growth. Similar to the initial run, the prediction was not good at the extremes. \Cref{tab:RFT1} shows the values of Mean Decrease Accuracy and Mean Decrease Gini sorted according to decreasing Mean Decrease Accuracy for the random forest model with 35 variables.

\begin{table}[]
\caption{Mean Decrease Accuracy (\%IncMSE) and Mean Decrease Gini (IncNodePurity) of each variable in the final run of  random forest}
\begin{tabular}{|l|l|l|lll}
\hline
Variable                             & \%IncMSE  & IncNodePurity & \multicolumn{1}{l|}{Variable}             & \multicolumn{1}{l|}{\%IncMSE} & \multicolumn{1}{l|}{IncNodePurity} \\ \hline
delta.K..MPa.sqrt.m..                & 21.331429 & 10.91146625   & \multicolumn{1}{l|}{Atm.Pressure}         & \multicolumn{1}{l|}{32.28836} & \multicolumn{1}{l|}{0.433617}      \\ \hline
log10.delta.K.                       & 20.689256 & 12.25333091   & \multicolumn{1}{l|}{R.ratio}              & \multicolumn{1}{l|}{9.326733} & \multicolumn{1}{l|}{0.497728}      \\ \hline
Temp..K.                             & 14.257944 & 4.18770728    & \multicolumn{1}{l|}{Thickness..mm.}       & \multicolumn{1}{l|}{10.59808} & \multicolumn{1}{l|}{0.57127}       \\ \hline
Min.Grain.Size..micro.m.             & 16.374741 & 0.22801076    & \multicolumn{1}{l|}{Yield.Strength..MPa.} & \multicolumn{1}{l|}{18.90627} & \multicolumn{1}{l|}{0.523596}      \\ \hline
Max.Grain.Size..micro.m.             & 12.057251 & 0.19262326    & \multicolumn{1}{l|}{Nickel..wt..}         & \multicolumn{1}{l|}{6.086034} & \multicolumn{1}{l|}{0.245061}      \\ \hline
Diff.in.GS.between.minor.major.phase & 11.389861 & 0.09075479    & \multicolumn{1}{l|}{Chromium..wt..}       & \multicolumn{1}{l|}{10.80819} & \multicolumn{1}{l|}{0.555393}      \\ \hline
HT1.Temp..K.                         & 12.560401 & 0.27023657    & \multicolumn{1}{l|}{Cobalt..wt..}         & \multicolumn{1}{l|}{5.340889} & \multicolumn{1}{l|}{0.602382}      \\ \hline
HT1.Time..hrs.                       & 3.993277  & 0.27102871    & \multicolumn{1}{l|}{Molybdenum..wt..}     & \multicolumn{1}{l|}{7.912807} & \multicolumn{1}{l|}{3.499861}      \\ \hline
HT1.CoolingRate..K.s.                & 4.325256  & 0.05469478    & \multicolumn{1}{l|}{Aluminium..wt..}      & \multicolumn{1}{l|}{9.512566} & \multicolumn{1}{l|}{0.091836}      \\ \hline
HT2.Temp..K.                         & 5.584551  & 0.74065123    & \multicolumn{1}{l|}{Titanium..wt..}       & \multicolumn{1}{l|}{5.118378} & \multicolumn{1}{l|}{0.446822}      \\ \hline
HT2.Time..hrs.                       & 5.299734  & 0.09739371    & \multicolumn{1}{l|}{Iron..wt..}           & \multicolumn{1}{l|}{7.332086} & \multicolumn{1}{l|}{0.094958}      \\ \hline
HT2.CoolingRate..K.s.                & 4.279956  & 0.06612501    & \multicolumn{1}{l|}{Carbon..wt..}         & \multicolumn{1}{l|}{5.975293} & \multicolumn{1}{l|}{0.3333}        \\ \hline
HT3.Temp..K.                         & 7.268819  & 0.09614288    & \multicolumn{1}{l|}{Boron..wt..}          & \multicolumn{1}{l|}{8.141424} & \multicolumn{1}{l|}{2.991938}      \\ \hline
HT3.Time..hrs.                       & 8.969581  & 0.17423834    & \multicolumn{1}{l|}{Zirconium..wt..}      & \multicolumn{1}{l|}{7.314942} & \multicolumn{1}{l|}{0.368474}      \\ \hline
Frequency..Hz.                       & 20.310028 & 14.9928676    & \multicolumn{1}{l|}{Silicon..wt..}        & \multicolumn{1}{l|}{8.056957} & \multicolumn{1}{l|}{0.137324}      \\ \hline
Loading.Time..s.                     & 11.324592 & 2.19791968    & \multicolumn{1}{l|}{Niobium..wt..}        & \multicolumn{1}{l|}{6.370777} & \multicolumn{1}{l|}{0.083549}      \\ \hline
Unloading.Time..s.                   & 7.346689  & 0.32059474    & \multicolumn{1}{l|}{Manganese..wt..}      & \multicolumn{1}{l|}{4.363146} & \multicolumn{1}{l|}{0.233896}      \\ \hline
Load.Shape                           & 7.432069  & 0.71008092    &                                           &                               &                                    \\ \cline{1-3}
\end{tabular}

\label{tab:RFT2}
\end{table}

\subsection{Neural Network}

\begin{figure}[htpb]
	\begin{center}
		\resizebox{140mm}{!}{\includegraphics{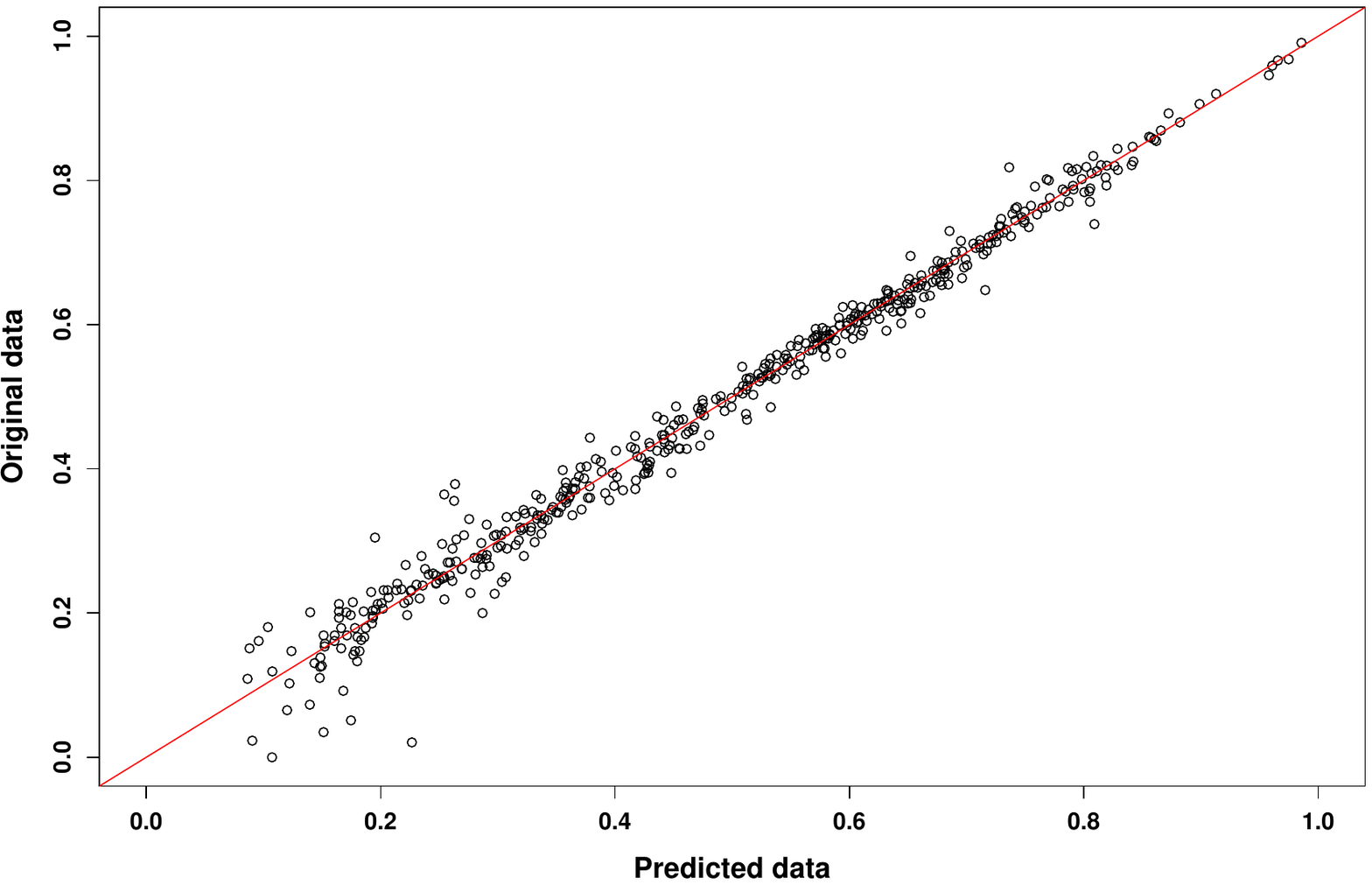}}
		\caption{Comparison of predicted $\log da/dN$ and the original data with neural network developed with one hidden layer. The red line represents $x=y$.}
		\label{fig:com_NN1}
	\end{center}
\end{figure}

\begin{figure}[htpb]
	\begin{center}
		\resizebox{140mm}{!}{\includegraphics{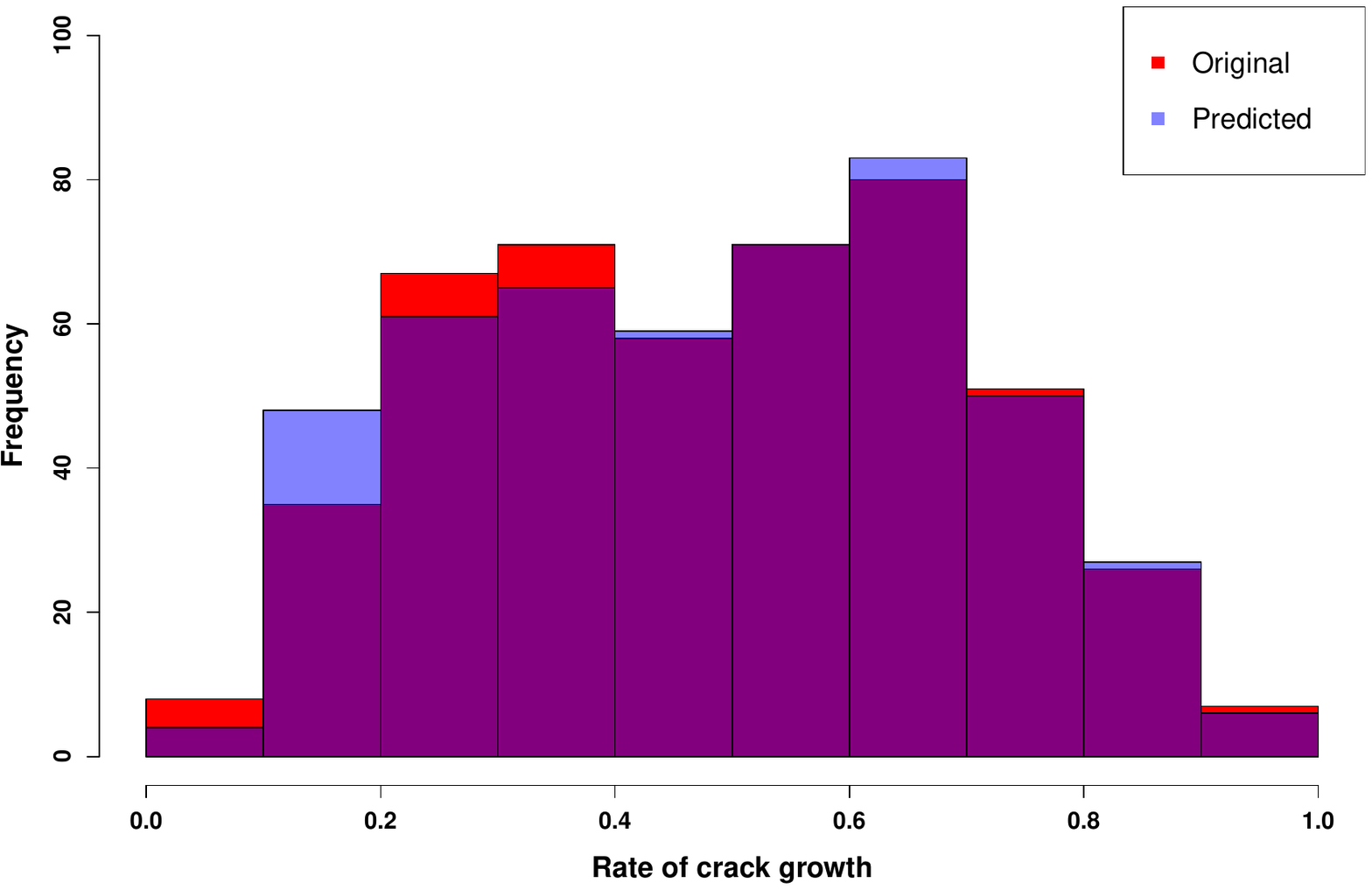}}
		\caption{Histogram comparing the distribution predicted and the original with neural network developed with one hidden layer.}
		\label{fig:hist_NN1}
	\end{center}
\end{figure}

For the first run with one hidden layer has the best $r^2$ value with 19 nodes in the hidden layer, when used for predicting the testing . The $r^2$ value for this framework is 0.9831. The correlation value of 0.9916 and p-value of $<2.2\times10^{-16}$ show there is a significant correlation with the chosen variables and the data. \Cref{fig:com_NN1} shows the comparison of the predicted values of $\log da/dN$ to the original values. It can be clearly seen from \Cref{fig:com_NN1} that the points align close to the $x=y$ line with major deviations present near the lower end similar to the initial run. This shows that the model is also not a good representation when the rate of crack growth is smaller, but slightly better than random forest. \Cref{fig:hist_NN1} shows the distribution of the original and the predicted rate of crack growth. It can be clearly observed that the distribution of the predicted data is much closer to the original data, which is better than what we observed in  the case of random forest.

For the second run with two hidden layers, the framework with 5 nodes in the second hidden layer has the best $r^2$ value of 0.9847 which is only 0.16\% better. This shows that adding another hidden layer does not significantly improve the performance of the neural net for being significantly computationally intensive for this data. So, a single hidden layer with 19 nodes is the best neural network framework for this data set.

\section{Conclusions}

In this work, the rate of fatigue crack growth in Nickel superalloys is estimated as a function of 51 available variables. Two different frameworks were used - random forest and neural network. Looking at the $r^2$ values of the predicted data using both the frameworks, the performance of neural network was only marginally better with 1.48\% improvement. It can be concluded that for Nickel superalloys, both random forest and neural network frameworks work well, but random forest is recommended in terms of computational speed. Neural network, although give a better performance, is much more computationally intensive and does not justify the marginal increase in accuracy. But in case, an explicit model is preferred to understand the relationship between the variables and the output, neural network is preferred.




%
% ...

\bibliographystyle{plainnat}

\bibliography{references}

\begin{thebibliography}{31}
\providecommand{\natexlab}[1]{#1}
\providecommand{\url}[1]{\texttt{#1}}
\expandafter\ifx\csname urlstyle\endcsname\relax
  \providecommand{\doi}[1]{doi: #1}\else
  \providecommand{\doi}{doi: \begingroup \urlstyle{rm}\Url}\fi

\bibitem[Bathias(1999)]{Bathias1999}
C.~Bathias.
\newblock {There is no infinite fatigue life in metallic materials}.
\newblock \emph{Fatigue and Fracture of Engineering Materials and Structures},
  22\penalty0 (7):\penalty0 559--565, jul 1999.
\newblock ISSN 8756758X.
\newblock \doi{10.1046/j.1460-2695.1999.00183.x}.

\bibitem[Bhadeshia(1999)]{Bhadeshia1999}
H.~K. D.~H. Bhadeshia.
\newblock {Neural Networks in Materials Science.}
\newblock \emph{ISIJ International}, 39\penalty0 (10):\penalty0 966--979, oct
  1999.
\newblock ISSN 0915-1559.
\newblock \doi{10.2355/isijinternational.39.966}.

\bibitem[Breiman(2002)]{breiman2002using}
L~Breiman.
\newblock Using models to infer mechanisms.
\newblock \emph{IMS Wald Lecture}, 2:\penalty0 59--71, 2002.

\bibitem[Breiman(1996)]{breiman1996out}
Leo Breiman.
\newblock Out-of-bag estimation.
\newblock 1996.

\bibitem[Butler et~al.(2018)Butler, Davies, Cartwright, Isayev, and
  Walsh]{butler2018machine}
Keith~T Butler, Daniel~W Davies, Hugh Cartwright, Olexandr Isayev, and Aron
  Walsh.
\newblock Machine learning for molecular and materials science.
\newblock \emph{Nature}, 559\penalty0 (7715):\penalty0 547--555, 2018.

\bibitem[Carrete et~al.(2014)Carrete, Li, Mingo, Wang, and
  Curtarolo]{Carrete2014}
Jes{\'{u}}s Carrete, Wu~Li, Natalio Mingo, Shidong Wang, and Stefano Curtarolo.
\newblock {Finding unprecedentedly low-thermal-conductivity half-heusler
  semiconductors via high-throughput materials modeling}.
\newblock \emph{Physical Review X}, 4\penalty0 (1):\penalty0 011019, feb 2014.
\newblock ISSN 21603308.
\newblock \doi{10.1103/PhysRevX.4.011019}.

\bibitem[Chen et~al.(2015)Chen, Chi, and Wang]{Chen2015}
Fu~Hsiang Chen, Der~Jang Chi, and Yi~Cheng Wang.
\newblock {Detecting biotechnology industry's earnings management using
  Bayesian network, principal component analysis, back propagation neural
  network, and decision tree}.
\newblock \emph{Economic Modelling}, 46:\penalty0 1--10, apr 2015.
\newblock ISSN 02649993.
\newblock \doi{10.1016/j.econmod.2014.12.035}.

\bibitem[{Fujll, H.; Mackay,D.J.C;
  Bhadeshia}(1996)]{FujllH.;MackayD.J.C;Bhadeshia1996}
H.K.D.H. {Fujll, H.; Mackay,D.J.C; Bhadeshia}.
\newblock {Network Growth Base}.
\newblock \emph{ISIJ International}, 36\penalty0 (1):\penalty0 1373--1382,
  1996.

\bibitem[Gislason et~al.(2006)Gislason, Benediktsson, and
  Sveinsson]{Gislason2006}
Pall~Oskar Gislason, Jon~Atli Benediktsson, and Johannes~R. Sveinsson.
\newblock {Random forests for land cover classification}.
\newblock In \emph{Pattern Recognition Letters}, volume~27, pages 294--300.
  North-Holland, mar 2006.
\newblock \doi{10.1016/j.patrec.2005.08.011}.

\bibitem[Grimm et~al.(2008)Grimm, Behrens, M{\"{a}}rker, and
  Elsenbeer]{Grimm2008}
R.~Grimm, T.~Behrens, M.~M{\"{a}}rker, and H.~Elsenbeer.
\newblock {Soil organic carbon concentrations and stocks on Barro Colorado
  Island - Digital soil mapping using Random Forests analysis}.
\newblock \emph{Geoderma}, 146\penalty0 (1-2):\penalty0 102--113, jul 2008.
\newblock ISSN 00167061.
\newblock \doi{10.1016/j.geoderma.2008.05.008}.

\bibitem[Ham et~al.(2005)Ham, Chen, Crawford, and Ghosh]{Ham2005}
Ji~Soo Ham, Yangchi Chen, Melba~M. Crawford, and Joydeep Ghosh.
\newblock {Investigation of the random forest framework for classification of
  hyperspectral data}.
\newblock In \emph{IEEE Transactions on Geoscience and Remote Sensing},
  volume~43, pages 492--501, mar 2005.
\newblock \doi{10.1109/TGRS.2004.842481}.

\bibitem[Hassan et~al.(2009)Hassan, Alrashdan, Hayajneh, and
  Mayyas]{hassan2009prediction}
Adel~Mahamood Hassan, Abdalla Alrashdan, Mohammed~T Hayajneh, and Ahmad~Turki
  Mayyas.
\newblock Prediction of density, porosity and hardness in
  aluminum--copper-based composite materials using artificial neural network.
\newblock \emph{Journal of materials processing technology}, 209\penalty0
  (2):\penalty0 894--899, 2009.

\bibitem[Karim et~al.(1997)Karim, Yoshida, Rivera, Saucedo, Eikens, and
  Gyu-Seop]{Karim1997}
M.~Nazmul Karim, Toshiomi Yoshida, Sheyla~L. Rivera, Victor~M. Saucedo,
  Bernhard Eikens, and O.~H. Gyu-Seop.
\newblock {Global and local neural network models in biotechnology: Application
  to different cultivation processes}, jan 1997.
\newblock ISSN 0922338X.

\bibitem[Kotkunde et~al.(2014)Kotkunde, Balu, Gupta, and Singh]{Kotkunde2014}
Nitin Kotkunde, Aditya Balu, Amit~Kumar Gupta, and Swadesh~Kumar Singh.
\newblock {Flow stress prediction of Ti-6Al-4V alloy at elevated temperature
  using artificial neural network}.
\newblock In \emph{Applied Mechanics and Materials}, volume 612, pages 83--88.
  Trans Tech Publications Ltd, 2014.
\newblock \doi{10.4028/www.scientific.net/AMM.612.83}.

\bibitem[Lawrence et~al.(2006)Lawrence, Wood, and Sheley]{Lawrence2006}
Rick~L. Lawrence, Shana~D. Wood, and Roger~L. Sheley.
\newblock {Mapping invasive plants using hyperspectral imagery and Breiman
  Cutler classifications (randomForest)}.
\newblock \emph{Remote Sensing of Environment}, 100\penalty0 (3):\penalty0
  356--362, feb 2006.
\newblock ISSN 00344257.
\newblock \doi{10.1016/j.rse.2005.10.014}.

\bibitem[Liaw and Wiener(2002)]{rrf}
Andy Liaw and Matthew Wiener.
\newblock Classification and regression by randomforest.
\newblock \emph{R News}, 2\penalty0 (3):\penalty0 18--22, 2002.

\bibitem[Montague and Morris(1994)]{Montague1994}
Gary Montague and Julian Morris.
\newblock {Neural-network contributions in biotechnology}, aug 1994.
\newblock ISSN 01677799.

\bibitem[Murakami et~al.(2000)Murakami, Nomoto, Ueda, and
  Murakami]{Murakami2000}
Y.~Murakami, T.~Nomoto, T.~Ueda, and Y.~Murakami.
\newblock {On the mechanism of fatigue failure in the superlong life regime
  (N{\textgreater}107 cycles). Part I: Influence of hydrogen trapped by
  inclusions}.
\newblock \emph{Fatigue and Fracture of Engineering Materials and Structures},
  23\penalty0 (11):\penalty0 893--902, nov 2000.
\newblock ISSN 8756758X.
\newblock \doi{10.1046/j.1460-2695.2000.00328.x}.

\bibitem[Nagasawa et~al.(2018)Nagasawa, Al-Naamani, and
  Saeki]{nagasawa2018computer}
Shinji Nagasawa, Eman Al-Naamani, and Akinori Saeki.
\newblock Computer-aided screening of conjugated polymers for organic solar
  cell: classification by random forest.
\newblock \emph{The journal of physical chemistry letters}, 9\penalty0
  (10):\penalty0 2639--2646, 2018.

\bibitem[Nakamoto et~al.(1990)Nakamoto, Fukunishi, and Moriizumi]{Nakamoto1990}
Takamichi Nakamoto, Katsufumi Fukunishi, and Toyosaka Moriizumi.
\newblock {Identification capability of odor sensor using quartz-resonator
  array and neural-network pattern recognition}.
\newblock \emph{Sensors and Actuators: B. Chemical}, 1\penalty0 (1-6):\penalty0
  473--476, jan 1990.
\newblock ISSN 09254005.
\newblock \doi{10.1016/0925-4005(90)80252-U}.

\bibitem[Pal(2005)]{Pal2005}
M.~Pal.
\newblock {Random forest classifier for remote sensing classification}.
\newblock \emph{International Journal of Remote Sensing}, 26\penalty0
  (1):\penalty0 217--222, jan 2005.
\newblock ISSN 0143-1161.
\newblock \doi{10.1080/01431160412331269698}.

\bibitem[Paris and Erdogan(1963)]{paris1963critical}
Pe~Paris and Fazil Erdogan.
\newblock A critical analysis of crack propagation laws.
\newblock 1963.

\bibitem[Parisi et~al.(1998)Parisi, {Di Claudio}, Lucarelli, and
  Orlandi]{Parisi1998}
R.~Parisi, E.~D. {Di Claudio}, G.~Lucarelli, and G.~Orlandi.
\newblock {Car plate recognition by neural networks and image processing}.
\newblock In \emph{Proceedings - IEEE International Symposium on Circuits and
  Systems}, volume~3, pages 195--198. IEEE, 1998.
\newblock \doi{10.1109/iscas.1998.703970}.

\bibitem[Pollock and Tin(2006)]{Pollock2006}
Tresa~M. Pollock and Sammy Tin.
\newblock {Nickel-based superalloys for advanced turbine engines: Chemistry,
  microstructure, and properties}, may 2006.
\newblock ISSN 15333876.

\bibitem[Poulton et~al.(1992)Poulton, Sternberg, and Glass]{Poulton1992}
Mary~M. Poulton, Ben~K. Sternberg, and Charles~E. Glass.
\newblock {Neural network pattern recognition of subsurface EM images}.
\newblock \emph{Journal of Applied Geophysics}, 29\penalty0 (1):\penalty0
  21--36, feb 1992.
\newblock ISSN 09269851.
\newblock \doi{10.1016/0926-9851(92)90010-I}.

\bibitem[Prasad et~al.(2006)Prasad, Iverson, and Liaw]{Prasad2006}
Anantha~M. Prasad, Louis~R. Iverson, and Andy Liaw.
\newblock {Newer classification and regression tree techniques: Bagging and
  random forests for ecological prediction}.
\newblock \emph{Ecosystems}, 9\penalty0 (2):\penalty0 181--199, mar 2006.
\newblock ISSN 14329840.
\newblock \doi{10.1007/s10021-005-0054-1}.

\bibitem[{R Core Team}(2017)]{R}
{R Core Team}.
\newblock \emph{R: A Language and Environment for Statistical Computing}.
\newblock R Foundation for Statistical Computing, Vienna, Austria, 2017.

\bibitem[Shyam et~al.(2004)Shyam, Torbet, Jha, Larsen, Caton, Szczepanski,
  Pollock, and Jones]{Shyam2004}
A~Shyam, C~J Torbet, S~K Jha, J~M Larsen, M~J Caton, C~J Szczepanski, T~M
  Pollock, and J~W Jones.
\newblock {DEVELOPMENT OF ULTRASONIC FATIGUE FOR RAPID, HIGH TEMPERATURE
  FATIGUE STUDIES IN TURBINE ENGINE MATERIALS}.
\newblock Technical report, 2004.

\bibitem[Singh et~al.(2011)Singh, Bhoopal, and Kumar]{singh2011prediction}
Ramvir Singh, RS~Bhoopal, and Sajjan Kumar.
\newblock Prediction of effective thermal conductivity of moist porous
  materials using artificial neural network approach.
\newblock \emph{Building and Environment}, 46\penalty0 (12):\penalty0
  2603--2608, 2011.

\bibitem[Suzuki et~al.(2006)Suzuki, Abe, MacMahon, and Doi]{Suzuki2006}
Kenji Suzuki, Hiroyuki Abe, Heber MacMahon, and Kunio Doi.
\newblock {Image-processing technique for suppressing ribs in chest radiographs
  by means of massive training artificial neural network (MTANN)}.
\newblock \emph{IEEE Transactions on Medical Imaging}, 25\penalty0
  (4):\penalty0 406--416, apr 2006.
\newblock ISSN 02780062.
\newblock \doi{10.1109/TMI.2006.871549}.

\bibitem[Vinci et~al.(2018)Vinci, Zoli, Sciti, Melandri, and
  Guicciardi]{vinci2018understanding}
Antonio Vinci, Luca Zoli, Diletta Sciti, Cesare Melandri, and Stefano
  Guicciardi.
\newblock Understanding the mechanical properties of novel uhtcmcs through
  random forest and regression tree analysis.
\newblock \emph{Materials \& Design}, 145:\penalty0 97--107, 2018.

\end{thebibliography}

\end{document}